\documentclass[final,5p,times,number]{elsarticle}
\usepackage{float}
\usepackage[caption=false,justification=justified]{subfig}  
\usepackage{graphicx}
\usepackage[dvipsnames]{xcolor}
\usepackage[english]{babel}
\usepackage{bm}
\usepackage{amsmath}
\usepackage{amssymb}
\usepackage{amsthm}
\usepackage{mathrsfs}
\usepackage[font=scriptsize]{caption}
\usepackage{ragged2e}
\usepackage[font=small,labelfont=bf,justification=raggedright]{caption}
\usepackage[utf8]{inputenc}
\usepackage[tikz]{bclogo}
\usepackage[framemethod=tikz]{mdframed}

\newsavebox{\measurebox}
\usepackage[colorlinks=false]{hyperref}

\begin{document}
	
\begin{frontmatter}
\title{On Noncommutative Quantum Mechanics and the Black-Scholes Model}
\author[IPN]{Abraham Espinoza-García\corref{cor1}}
\cortext[cor1]{Corresponding author.}\ead{aespinoza@ipn.mx}
\author[IPN,IESFi]{Pablo Vega-Lara}\ead{pvegal@ipn.mx}
\author[IPN]{Luis Rey Díaz-Barrón}\ead{lrdiaz@ipn.mx}
\author[IPN,IESFi]{F. Teodoro Hernández Grovas}\ead{fhernandezg@ipn.mx}

\affiliation[IPN]
{organization={Unidad Profesional Interdisciplinaria de Ingeniería Campus Guanajuato del Instituto Politécnico Nacional},
addressline={Av. Mineral de Valenciana No. 200, Col. Fraccionamiento Industrial Puerto Interior}, 
city={Silao de la Victoria},
postcode={36275}, 
state={Guanajuato},
country={México}}
\affiliation[IESFi]
{organization={Instituto de Estudios Superiores en Finanzas (IESFi)},
addressline={Blvd. García de León 1035 piso 4, Col. Chapultepec Sur}, 
city={Morelia},
postcode={58260}, 
state={Michoacán},
country={México}}		
\begin{abstract}
Two novel and direct quantum mechanical representations of the Black-Scholes model are constructed based on the (Wick-rotated) quantization of two specific mechanical systems. The quantum setup is achieved by means of the associated Laplace-Beltrami operator (one for each model), and not by merely applying the naive rule $p\mapsto-i\hbar\partial_q$. Additionally, the clear identification of the geometric content of the underlying classical framework is exploited in order to arrive at a noncommutative quantum mechanics generalization of the Black-Scholes model. We also consider a system consisting of two degrees of freedom whose (Wick-rotated) quantization leads to a model which can be seen as related to the Merton-Garman family. This model is also generalized via noncommutative quantum mechanics.
\end{abstract}
		
		
\end{frontmatter}

\section{Introduction}

Here we recall the quantum mechanical representation of the Black-Scholes model.

\subsection{The Black-Scholes formula}
A mathematical model for Brownian motion and its use in the description of financial instruments was first put forward by Bachelier \cite{bachelier} (Einstein provided an independent description of Brownian motion five years later \cite{einstein}).

The so called Ito calculus (in particular, the celebrated Ito formula) \cite{ito} is the cornerstone of the deterministic Black-Scholes (BS) model for option pricing. The Ito stochastic differential equation (SDE) is of the form\cite{ito},\cite{paul}
$$dx=a(x(t),t)dt+b(x(t),t)dW(t),$$
where $W$ is a Wiener process ($(dW)^2=dt$)\cite{wiener},\cite{larnold}. 

Ito's formula for $f(x(t),t)$ (where $x$ obeys the above Ito process) can be heuristically expressed in terms of differentials as\cite{paul}, (for a more rigorous proof see, e.g. \cite{larnold})
\begin{equation}
df=\left[f_t+a(x(t),t)f_x+\frac{1}{2}b^2(x(t),t)f_{xx}\right]dt+b(x(t),t)dW\label{ito-formula}
\end{equation}

The particular Ito SDE
\begin{equation}
dS=\mu S(t)dt+\sigma S(t)dW(t),\label{brownian-stock}
\end{equation}
models (as a geometric Brownian motion) the evolution of stock prices on an efficient market (for the relevant finance-related concepts, see e.g. \cite{hull}), with $\mu$ and $\sigma$ constants ($\sigma$ is identified with the so called volatility of the stock price) \cite{paul}.  

Ito's formula for a \textit{derivative} $C(S(t),t)$ of $S(t)$ (e.g. an \textit{option}) gives
\begin{align}
dC&=\left[C_t+\mu S(t)C_S+\frac{1}{2}\left(\sigma S(t)\right)^2C_{SS}\right]dt+\sigma S(t)C_SdW\nonumber\\
  &=\left[C_t+\frac{1}{2}\left(\sigma S(t)\right)^2C_{SS}\right]dt+C_SdS\label{ito-formula-C}
\end{align}
where in the last step we used \eqref{brownian-stock}.

Now, consider a perfectly hedged portfolio 
\begin{equation}
\Pi(t)=C(S(t),t)-\epsilon S(t),\label{portfolio}
\end{equation}
i.e. $d\Pi=r\Pi dt$ ($r$ the risk-free interest rate), with $\epsilon$ a parameter taking values on $[0,1]$ (``delta hedging''). From the definition \eqref{portfolio} of the portfolio $\Pi$ it follows that 
$$dC=d\Pi+\epsilon dS=r\Pi dt+\epsilon dS$$ 
comparing with \eqref{ito-formula-C} we must have,
\begin{equation}
r\Pi=C_t+\frac{1}{2}\left(\sigma S(t)\right)^2C_{SS},\quad\epsilon=C_S.
\end{equation}
Using again \eqref{portfolio} we finally obtain,
\begin{equation}
rC-rSC_S=C_t+\frac{1}{2}\left(\sigma S(t)\right)^2C_{SS},\label{black-scholes}
\end{equation}
which is the celebrated \textit{Black-Scholes equation} (for european options).

\subsection{Black-Scholes equation as a Schr\"odinger-like equation}

Two main avenues have been explored in order to arrive at a quantum representation of the BS model. In the first one, by performing several change of variables, the BS equation \eqref{black-scholes} is first rendered as a heat equation (see, e.g. \cite{paul}), 
and then the well-known mapping between a diffusion-type equation and (a Wick-rotated) Schr\"odinger equation is applied. Specifically, by considering the variable $S=e^x$ the BS equation \eqref{black-scholes} takes the form
\begin{equation}
\frac{\partial C}{\partial t} + \frac{1}{2}\sigma^2\frac{\partial^2 C}{\partial x^2}+\left[r-\frac{1}{2}\sigma^2\right]\frac{\partial C}{\partial x} =rC
\end{equation}
By implementing further 
\begin{align*}
C(x,t)=\exp{\left[\frac{1}{\sigma^2}\left(\frac{\sigma^2}{2}-r\right)x+\frac{1}{2\sigma^2}\left(\frac{\sigma^2}{2}+r\right)^2 t\right]} \psi(x,t);\\
\tau=T-t,
\end{align*}
the heat equation
\begin{equation}
\frac{\partial \psi}{\partial \tau}=\frac{\sigma^2}{2}\frac{\partial^2\psi}{\partial x^2}
\end{equation}
follows. A somewhat prominent investigation using this particular path is \cite{contreras}. This first route is deeply ingrained in the interdisciplinary field of econophysics, in which techniques of theoretical physics---notably, statistical mechanics---are  utilized in order to study empirical and theoretical aspects of economics and finance (a well known reference is \cite{mantegna-stanley}).

In the second approach to a quantum mechanical description of the BS model, one arrives to a Schr\"odinger-like equation in which a velocity-dependent term is featured (see, e.g. \cite{baaquie1}-\cite{haven}). In this proposal, it is assumed that the BS formula \eqref{black-scholes} written in the new variable $S=e^x$ can alternatively be obtained by the naive quantization prescription $p\mapsto-i\hbar\partial_q$ (and further Wick-rotation) of a classical mechanical system whose Hamiltonian is
\begin{equation}
\tilde{\mathcal{H}}(x,p)=\frac{\sigma^2}{2}p^2+\left(\frac{\sigma^2}{2}-r\right)p+r,\label{baaquie-class-ham}
\end{equation}
where a term linear in the momentum $p$ (canonically conjugated to the configuration variable $x$) is featured.
%

We stress that, in both types of proposals, a quantum mechanical interpretation of the BS model is sought only after rewriting the BS equation as an easily-recognized Schr\"odinger-like equation.


The paper is displayed as follows. The second section is devoted to establish that the BS model is already an ``out of the box'' (Wick-rotated) quantum mechanical system, and to highlight the geometric content carried by its classical limit. In Section 3 we succinctly recall the particular flavor of noncommutative quantum mechanics which we will be focusing on in order to carry the noncommutative generalization of the BS model. In Section 4 the Black-Scholes model is generalized by means of the noncommutative quantum mechanics paradigm, based on the framework put forward in Sections 2 and 3. A system consisting of two degrees of freedom is considered in section 5, which is shown to be closely related to a subset of the Merton-Garman family of models. Two noncommutative generalizations are provided for this model.
In the final section 6 we spell out what we consider to be the most important attributes of our investigation. We provide there a kind of summary---with significant equations reproduced for the benefit of the eager reader---and state possible trends which can be followed in order to pursue complimentary studies within the generalized geometric framework presented in the main parts of the present document. 
\section{Direct quantum mechanical representation of the Black-Scholes model and its classical limit}\label{BS}

Let $(\mathcal Q,g)$ be a configuration space manifold (with metric field $g$) of some classical mechanical system, locally coordinatized by $q=(q^1,\dots,q^n)$, and with associated Lagrangian (small letters run from $1$ to $n=\dim(\mathcal Q)$)
\begin{equation}
\mathcal{L}(q,\dot{q})=\frac{1}{2}g_{ab}\dot{q}^{a}\dot{q}^{b}-U(q).
\end{equation}
The Legendre mapping 
\begin{equation}
p_a=\frac{\partial\mathcal{L}}{\partial \dot{q}^a}
\end{equation}
leads to the associated Hamiltonian function
\begin{equation}
\mathcal{H}(q,p)=\frac{1}{2}g^{ab}p_{a}p_{b}+U(q)=K(q,p)+U(q),\label{class-ham}
\end{equation}
where the mass of the particle has been obviated.

Recall that, from the geometric point of view, the phase space associated to a general configuration space manifold $(\mathcal Q,g)$ is the symplectic manifold $(\Gamma, \omega)$, where $\Gamma$ is the cotangent bundle $T^{*}\mathcal{Q}$ of $\mathcal{Q}$ and $\omega$ is the (canonical) symplectic $2$-form on $T^{*}\mathcal{Q}$ (for detalis, see, e.g. \cite{marsden}-\cite{arnold}). The inverse $\pi$ of $\omega$ (we follow the convention used in \cite{novikov}), i.e. 
\begin{equation}
\pi^{AC}\omega_{CB}=\delta^{A}_{B}, \label{poisson-symplectic-relation}
\end{equation}
is the so called Poisson structure, which defines the Hamiltonian vector field $H$ by (capital letters run from $1$ to $N=2n=\dim(\Gamma)$)
\begin{equation}
H^A=\pi^{AB}\partial_{B}\mathcal H.
\end{equation}
This vector field is very important since its flow gives the classical trajectories followed by the mechanical system (i.e, it gives the solution curves to the equations of motion). Specifically, Hamilton's equations can be written, in \emph{any coordinate patch} of $(\Gamma,\omega)$, as 
\begin{equation}
\dot{\zeta}^{A}=H^A(\zeta),
\end{equation}
where $\zeta=(\zeta^1,\dots,\zeta^{N})$ are local coordinates associated to the patch, in which (for canonical coordinates) the first $n$ are to be interpreted as configuration variables, whereas the remaining $n$ are the momenta variables.

Poisson brackets on $(\Gamma,\omega)$ are defined by 
\begin{equation}
\{f,g\}=\pi^{AB}\partial_Af\partial_Bg.\label{poisson-structure}
\end{equation}
Due to the Darboux theorem of symplectic geometry, there always exists a local coordinate system $x=(q^1,\dots,q^n,p_1,\dots,p_n)$ for an open region $U\subseteq\Gamma$ such that $\omega$ can be written as 
\begin{align}
\omega_x&=dx^{n+1}\wedge dx^{1}+dx^{n+2}\wedge dx^{2}+\dots+dx^{2n}\wedge dx^{n}\nonumber\\
&=dp_a\wedge dq^a.
\end{align}

Phase space coordinates satisfying the above requirement are usually called canonical, and the local chart they define is termed symplectic. 
Hence, in canonical coordinates $\pi$ acquires the form
\begin{align}
\pi_x&=\partial_{1}\wedge\partial_{n+1}+\partial_{2}\wedge\partial_{n+2}+\dots+\partial_{n}\wedge\partial_{2n}\nonumber\\
&=\partial_{q^a}\wedge\partial_{p_a}.
\end{align}

Recall now that the canonical quantization of the above mechanical system, in the standard Schr\"odinger representation, is achieved by setting (see, e.g. \cite{greiner}, \cite{marsden}) 
\begin{align}
\mathcal{H}=K(q,p)+U(q)\mapsto\hat{\mathcal{H}}=-\frac{\hbar^2}{2}\Delta_{q}+U(q),\\ 
\Delta_{q}:=\frac{1}{\sqrt{|\det(g)|}}\partial_a\left[\left(\sqrt{|\det(g)|}g^{ab}\right)\partial_b\right],\label{laplace-beltrami}
\end{align}
where the differential operator $\Delta_q$ is sometimes referred to as the Laplace-Beltrami operator (in coordinates $q$). This prescription reduces to the usual one for the case in which $\mathcal Q$ is flat and $q$ are the associated flat (cartesian) coordinates.

If the BS model is to be given a proper (Wick-rotated) quantum mechanical interpretation, the non-temporal part of the differential operator featured on the BS equation must be matched to the Laplace-Beltrami operator via the above prescription. As a by product, the whole geometric structure of the associated classical limit would be obtained, therefore making available well-known global geometric techniques usually employed at the (semi-)classical level in modern theoretical physics.
\subsection{Revisiting the first approach for a quantum-mechanical representation of the BS model}
Consider therefore the one-dimensional configuration manifold  $(\mathcal{Q}^{BS},g)$ coordinatized by $q^1=q>0$ and where $g^{11}=q^2$, with Lagrangian
\begin{equation}
\mathcal{L}^{BS}(q,\dot{q})=\frac{1}{2}\left(\frac{\dot{q}}{q}\right)^2-U(q).\label{lag-1}
\end{equation}
The corresponding Hamiltonian function is hence given by
\begin{equation}
\mathcal{H}^{BS}(q,p)=\frac{1}{2}q^2p^2+U(q)=K^{BS}(q,p)+U(q).\label{class-ham1a}
\end{equation}
Since the momentum $p$ and Hamiltonian $\mathcal{H}^{BS}$ were constructed according to the standard prescription given by the Legendre transform, the phase space patch defined by $x=(q,p)$ is automatically a symplectic one, and so the symplectic structure $\omega^{BS}$ acquires its flat form in these coordinates: $\omega^{BS}_{x}=dp\wedge dq$.

Now, the associated Laplace-Beltrami operator is 
\begin{equation}
K^{BS}(q,p)\mapsto-\frac{\hbar^2}{2}\Delta^{BS}_q=-\frac{\hbar^2}{2}\left(q^2\partial^2_q+q\partial_q\right)\label{laplace-beltrami-BS} 
\end{equation}
and so, the quantized model is defined by the Schr\"odinger equation
\begin{equation}
i\partial_t\psi(q,t)=\left[-\frac{\hbar^2}{2}\left(q^2\partial^2_q+q\partial_q\right)+U(q)\right]\psi(q,t).\label{schr-BS1a}
\end{equation}
A direct comparison with tne BS formula \eqref{black-scholes} allows (upon fixing $U(q)$) for an exact Wick-rotated (i.e. $t\mapsto-it$) quantum mechanical interpretation of the BS model, in which $S$ takes the role of $q$ (while $\sigma^{2}$ could be seen as the analog of the reciprocal of the mass of the particle). Note that in this more direct quantum mechanical representation of the BS formula it is mandatory that $r$ be proportional to $\sigma^2$. 

Let us now seek a simpler classical description in terms of another set of canonical coordinates. Take the local chart defined by $\bar{x}=(\bar{q}=\ln(q),\bar{p}=qp)$, on the BS classical mechanical system $(\Gamma^{BS},\mathcal{H}^{BS})$. It is straight-forward to verify that the associated local trivialization is indeed a symplectic one.
The Hamiltonian \eqref{class-ham1a} is written in these coordinates as,
\begin{equation}
\mathcal{H}^{BS}(\bar{q},\bar{p})=\frac{1}{2}\bar{p}^2+U(q)=K^{BS}(\bar{p})+U(\bar{q}).\label{class-ham1b}
\end{equation}
Not surprsingly, the Laplace-Beltrami operator takes the standard flat form, 
\begin{equation}
K^{BS}(\bar{p})\mapsto-\frac{\hbar^2}{2}\Delta^{BS}_{\bar{q}}=-\frac{\hbar^2}{2}\partial^2_{\bar{q}},
\end{equation}
and so, the quantum model is defined by the standard Schr\"odinger equation
\begin{equation}
i\partial_t\psi(\bar{q},t)=\left[-\frac{\hbar^2}{2}\partial^2_{\bar{q}}+U(\bar{q})\right]\psi(\bar{q},t).\label{schr-B1b}
\end{equation}
Therefore, in such new canonical coordinates $(\bar{q},\bar{p})$, (and with the identifications spelled out before) the BS model acquires the Wick-rotated quantum mechanical interpretation referred to as the first avenue in the introductory section.

\subsection{Revisiting the second approach for a quantum-mechanical representation of the BS model}

As with the first avenue, let us try to accommodate this second proposal along the lines of our more well grounded and direct method. Firstly, consider a classical system described by the same configuration manifold as before, $(\mathcal Q^{BS},g)$, $g^{11}=q^2$, but this time with a velocity-dependent Lagrangian 
\begin{equation}
\tilde{\mathcal{L}}^{BS}(q,\dot{q})=\frac{1}{2}\left(\frac{\dot q}{q}\right)^2-U(q)+\alpha\frac{\dot{q}}{q}\label{lag-2}
\end{equation}
the associated Hamiltonian is hence
\begin{equation}
\tilde{\mathcal{H}}^{BS}(q,p)=\frac{1}{2}q^2p^2-\alpha qp+U(q)+\frac{\alpha^2}{2}=\tilde{K}^{BS}(q,p)+\tilde{U}(q,p),\label{class-ham2a}
\end{equation}
where
\begin{equation}
\tilde{K}^{BS}(q,p)=\frac{1}{2}q^2p^2\quad\mathrm{and}\quad\tilde{U}(q,p)=-\alpha qp+U(q)+\frac{\alpha^2}{2}
\end{equation}
are, by construction, the corresponding kinetic and potential terms, respectively. We stress here that the symplectic structure of the classical system $(\tilde{\Gamma}^{BS},\tilde{\mathcal{H}}^{BS})$ is also $\omega^{BS}$---the same as that of $(\Gamma^{BS},\mathcal{H}^{BS})$.

Upon quantization, apart from promoting 
\begin{equation}
\tilde{K}^{BS}(q,p)\mapsto-\frac{\hbar^2}{2}{\Delta}^{BS}_q=-\frac{\hbar^2}{2}\left(q^2\partial_q^2+q\partial_q\right),
\end{equation}
we must also take care of the velocity-dependent term in the potential $\tilde{U}(q,p)$. In this case, three factor orderings are readily envisaged, namely $qp$, $pq$ and $\frac{1}{2}(qp+pq)$. These would produce different Schr\"odinger equations. Keeping with the simplest ordering ($qp$), we obtain the following Schr\"odinger equation, 
\begin{align}
i\partial_t\psi(q,t)=\left[-\frac{\hbar^2}{2}\left(q^2\partial_q^2+q\partial_q\right)-\alpha q(-i\hbar\partial_q)\right.\nonumber\\
\left.+U(q)+\frac{\alpha^2}{2}\right]\psi(q,t),\label{schr-BS-2a}
\end{align}
where we have used the usual recipe, $p\mapsto-i\hbar\partial_q$, \emph{only in promoting the velocity-dependent term} to a differential operator. The Wick-rotated version of the above equation can be made to match the BS formula by fixing $U(q)$ and the parameter $\alpha$.

We consider now the coordinates $(\bar{q},\bar{p})$ used before in the first case. In these canonical coordinates, the classical Hamiltonian \eqref{class-ham2a} is written as
\begin{equation}
\tilde{\mathcal{H}}^{BS}(\bar{q},\bar{p})=\frac{1}{2}\bar{p}^2-\alpha\bar{p}+U(\bar{q})+\frac{1}{2}\alpha^2=\tilde{K}^{BS}(\bar{p})+\tilde{U}(\bar{q},\bar{p}),\label{class-ham2b} 
\end{equation}
where
\begin{equation}
\tilde{K}^{BS}(\bar{q},\bar{p})=\frac{1}{2}\bar{p}^2\quad\mathrm{and}\quad\tilde{U}(\bar{q},\bar{p})=-\alpha\bar{p}+U(\bar{q})+\frac{\alpha^2}{2}.
\end{equation}
The quantization of the model can now be carried out more straight-forwardly (as was the case with \eqref{class-ham1b}). As before, apart from promoting 
\begin{equation}
\tilde{K}^{BS}(\bar{p})\mapsto-\frac{\hbar^2}{2}{\Delta}^{BS}_{\bar{q}}=-\frac{\hbar^2}{2}\partial_{\bar{q}}^2,
\end{equation}
we need to take care of the velocity-dependent term featured in the potential $\tilde{U}(\bar{q},\bar{p})$. The corresponding Schr\"odinger equation is
\begin{align}
i\partial_t\psi(\bar{q},t)=\left[-\frac{\hbar^2}{2}\partial_{\bar q}^2-\alpha(-i\hbar\partial_{\bar{q}})
+U(\bar{q})+\frac{\alpha^2}{2}\right]\psi(\bar{q},t),\label{schr-BS2}
\end{align}
where we have again used the usual prescription $\bar{p}\mapsto-i\hbar\partial_{\bar{q}}$ only to tackle the velocity-dependent term in the potential. The Wick-rotated version of the above equation can be matched to the BS formula by fixing $U(\bar{q})$ and the parameter $\alpha$.

It is instructive to note that, upon taking $\alpha\to0$, both classical and quantum scenarios defined by \eqref{lag-2} go over to those associated to \eqref{lag-1}. 

We have therefore managed to give a more proper treatment (from a theoretical physics perspective) to the two more commonly used quantum-mechanical representations of the BS model. The insight gained will prove to be crucial in seeking an appropriate extension of the BS model to the realm of the so called noncommutative quantum mechanics.

\section{Brief account on some of the ideas of noncommutative quantum mechanics and its classical limit}\label{ncqm}

Here we give a very succinct account on the somewhat contemporary proposal known as noncommutative quantum mechanics. A short but useful review can be found in Ref. \cite{ncqm-review}.

In simple terms, the philosophy of noncommutative quantum mechanics is that of generalizing the standard quantum mechanics commutation relations 
\begin{equation}
\left[\hat{q}^a,\hat{p}_b\right]=i\hbar\delta^a_b\hat{1},\quad\left[\hat{q}^a,\hat{q}^b\right]=\hat{0}=\left[\hat{p}_a,\hat{p}_b\right]\label{can-q-algebra}
\end{equation}
in order to allow for additional dispersion relations. The motivation relies mainly on the fact that prominent candidates for a theory of quantum gravity predict that spacetime has a sort of discrete nature at a fundamental level (see, for instance, \cite{sabine}). Indeed, particular limits of string/M theory lead to what is now called noncommutative gauge theories (see, e.g. \cite{nekrasov} and \cite{szabo}); also, in loop quantum gravity, we have noncommutativity of so called fluxes, and a discrete spectrum associated to fundamental spatial operators \cite{lqg-review}. Noncommutative quantum mechanics is a simplified framework in which such kinds of discretizations can be accounted for in an effective way. Actually, the finite-degrees-of-freedom limit of noncommutative gauge theories leads naturally to different incarnations of noncommutative quantum mechanics.

For the moment we focus on a particularly simple noncommutative structure, which will serve as a preliminary example in order to tackle more involved noncommutativity schemes. Consider a quantum mechanical system defined by a certain Hamiltonian operator $\hat{H}(\hat{p},\hat{q})$ but in which the slightly more general quantum algebra
\begin{equation}
\left[\hat{q}^a,\hat{p}_b\right]=i\hbar\delta^{a}_{b}\hat{1}\left(1+\theta\right)\,\quad\left[\hat{q}^a,\hat{q}^b\right]=\hat{0}=\left[\hat{p}_a,\hat{p}_b\right]\label{nc-q-algebra}
\end{equation}
is to be satisfied, where $\theta$ is a constant real parameter. This would entail a correction to the standard quantum mechanics dispersion relations for the simultaneous measurement of so called observables. Ordinary quantum mechanics would be recovered by taking $\theta\to0$. 

Now, assuming the Dirac correspondence between Poisson brackets and commutators to be still valid in this enlarged setup, the classical counterpart to the above commutation relations is 
\begin{equation}
\left\{q^a,p_b\right\}=\delta^a_b(1+\theta),\quad\left\{q^a,q^b\right\}=0=\left\{p_a,p_b\right\}\label{nc-c-algebra}.
\end{equation}
This in turn means, according to \eqref{poisson-structure}, that the Poisson structure of the classical counterpart to the noncommutative quantum model is given (in coordinates $x=(q^1,\dots,q^n,p_1,\dots,p_n)$) by
\begin{equation}
{\pi_\theta}_x=\left(1+\theta\right)\partial_{q^a}\wedge\partial_{p_a},\label{nc-pois-struc}
\end{equation}
The corresponding symplectic structure (according to \eqref{poisson-symplectic-relation}) is therefore expressed in coordinates $x$ as
\begin{equation}
{\omega_\theta}_x=\left(1+\theta\right)^{-1}dp_a\wedge dq^a,\label{nc-symp-struc}
\end{equation}
and so, the chart defined by coordinates $x$ is not exactly symplectic. Therefore, the noncommutative quantum model defined by $\hat{\mathcal{H}}\left(\hat{q},\hat{p}\right)$ and \eqref{nc-q-algebra} can be considered as the quantization of the classical system $(\Gamma_\theta,\mathcal{H})$, where $\omega_\theta$ is the symplectic form on the classical phase space $\Gamma_\theta$ (which is itself the cotangent bundle of some underlying configuration manifold $(\mathcal{Q}_\theta,g_\theta)$). 

By going to canonical coordinates $\underline{x}=\left(\underline{q}^1,\dots,\underline{q}^n,\underline{p}_1,\dots,\underline{p}_n\right)$ and proceeding to quantization (via the corresponding Laplace-Beltrami operator), a standard canonical algebra of the form \eqref{can-q-algebra} and a corresponding Schr\"odinger-like equation would be obtained, therefore representing the noncommutative quantum model in a way akin to the standard commutative one.

We have therefore the following situation. Consider a classical mechanical system defined by the pair $(\Gamma,\mathcal{H})$ where $\omega_x=dp_a\wedge dq^a$ is the symplectic structure on $\Gamma$ (so that the coordinates $x=(q^1,\dots,q^n,p_1,\dots,p_n)$ define a symplectic chart for some open region of $\Gamma$ is defined by ). Let $(\Gamma_\theta,\mathcal{H})$ be another mechanical system with the same Hamiltonian function $\mathcal{H}$ as the former, and where $\omega_\theta$ is given by \eqref{nc-symp-struc} (reducing to $\omega$ for $\theta\to0$). Hence the system $(\Gamma_\theta,\mathcal H)$ reduces to the mechanical system $(\Gamma,\mathcal{H})$ for $\theta\to 0$, and so it is appropriately called a noncommutative generalization of system $(\Gamma,\mathcal H)$. If noncommutative quantum mechanics indeed results to be a genuine correction to standard quantum mechanics, such correction would descend to the classical level selecting $(\Gamma_\theta,\mathcal H)$ as the correct classical model (with $(\Gamma,\mathcal H)$ serving only as a preliminary classical setup). The quantization of the second system $(\Gamma_\theta,\mathcal{H})$ leads therefore to a new quantum model $(\hat\Gamma_\theta,\hat{\mathcal{H}})$ (reducing itself to the one obtained from the quantization of $(\Gamma,\mathcal H)$, which we denote by $(\hat{\Gamma},\hat{\mathcal H})$). This new quantum model $(\hat\Gamma_\theta,\hat{\mathcal{H}})$ is the noncommutative quantum mechanics generalization of $(\hat{\Gamma},\hat{\mathcal H})$. 

The noncommutative quantum mechanics defined by the generalized relations \eqref{nc-q-algebra} is of a simple form in the sense that $\theta$ is assumed to be constant in the coordinate patch defined by $x$. We could of course do away with this restriction and consider a general dependency $\theta(x)$. 
\section{Noncommutative generalization of the Black-Scholes model}
%

In both of the \emph{original models} $(\Gamma^{BS},\mathcal{H}^{BS})$ and $(\tilde{\Gamma}^{BS},\tilde{\mathcal{H}}^{BS})$, the only non-trivial relation is 
\begin{equation}
\{q,p\}=1\longleftrightarrow [\hat{q},\hat{p}]=i\hbar\hat{1}.
\end{equation}
We first work in the phase space patch defined by the ``initial'' coordinates $x=(q,p)$ (which, as we saw, directly lead to the BS equation via \eqref{lag-1}).

A general noncommutative extension is therefore of the form 
\begin{equation}
\{q,p\}=1+\theta F(q,p)\longleftrightarrow [\hat{q},\hat{p}]=i\hbar\left(\hat{1}+\theta\hat{F}(\hat{q},\hat{p})\right).\label{nc-general-algebra}
\end{equation}
Recall that, from the discussion given in Section \ref{ncqm}, we can establish that the above relations define classical systems $(\Gamma_\theta^{BS},\mathcal{H}^{BS})$ and $(\tilde{\Gamma}_\theta^{BS},\tilde{\mathcal{H}}^{BS})$ such that
\begin{itemize}
\item their Hamiltonians are the same as the ones featured in the original commutative models (which are, \eqref{class-ham1a} and \eqref{class-ham2a}) when written in the coordinate patch defined by $x=(q,p)$; 
\item their common symplectic structure $\omega^{BS}_\theta$ is \emph{not} in canonical form when expressed in coordinates $x=(q,p)$; 
\item upon taking $\theta\to0$, they reduce to $(\Gamma^{BS},\mathcal{H}^{BS})$ and $(\tilde{\Gamma}^{BS},\tilde{\mathcal{H}}^{BS})$, respectively; 
\item the corresponding quantum versions reduce (for $\theta\to0$) to the standard quantum models defined by \eqref{schr-BS1a} and \eqref{schr-BS2}, respectively.
\end{itemize}
We emphasize that the fundamental classical framework developed in Section \ref{BS} is of paramount importance for the construction of the noncommutative setup depicted above.

In the following, we specialize the general proposal \eqref{nc-general-algebra} to 
\begin{equation}
\{q,p\}=1+\theta f(q)\longleftrightarrow [\hat{q},\hat{p}]=i\hbar\left(\hat{1}+\theta\hat{f}(\hat{q})\right).\label{nc-algebra}
\end{equation}
Thus, according to \eqref{nc-algebra}, the Poisson structure $\pi_\theta^{BS}$ associated to both deformed classical models $(\Gamma_\theta^{BS},\mathcal{H}^{BS})$ and $(\tilde{\Gamma}_\theta^{BS},\tilde{\mathcal{H}}^{BS})$ is given (in coordinates $x$) by
\begin{equation}
{\pi_{\theta}^{BS}}_{x}=\left(1+\theta f(q)\right)\partial_{q}\wedge\partial_{p}.\label{def-pois-structure}
\end{equation}
Hence, the corresponding symplectic structure $\omega_\theta^{BS}$ can be written as   
\begin{equation}
{\omega_{\theta}^{BS}}_{x}=\left(1+\theta f(q)\right)^{-1}dp\wedge dq.\label{def-symp-structure}
\end{equation}

\subsection{Noncommtuative extension based on the first approach for a quantum-mechanical representation of the BS model}
In order to arrive at more standard classical and quantum formulations for the correspondence \eqref{nc-algebra}, we  seek coordinates $\underline{x}=(\underline{q},\underline{p})$ in which the symplectic structure \eqref{def-symp-structure} can be written in flat form, i.e., 
\begin{equation}
{\omega^{BS}_{\theta}}_{\underline{x}}=d\underline{p}\wedge d\underline{q}.
\end{equation}
One can easily verify that one such patch is defined by the relations
\begin{equation}
q=\underline{q},\quad p=\underline{p}+\theta\underline{p}f(\underline{q}).
\end{equation}
The Hamiltonian \eqref{class-ham1a} is written in these coordinates as
\begin{equation}
\mathcal{H}^{BS}\left(\underline{q},\underline{p}\right)=\frac{1}{2}\underline{q}^2\left[\underline{p}+\theta\underline{p} f\left(\underline{q}\right)\right]^2+U\left(\underline{q}\right)=K^{BS}_\theta\left(\underline{q},\underline{p}\right)+U\left(\underline{q}\right)
\end{equation}
Note that in this case 
\begin{equation}
{g_\theta}^{11}=\left(\underline{q}+\theta\underline{q}f\left(\underline{q}\right)\right)^2\label{metric-nc-BS} 
\end{equation}
defines the metric on the underlying configuration manifold $(\mathcal{Q}^{BS}_\theta,g_\theta)$.

Now that we have our noncommutative classical framework $(\Gamma^{BS}_{\theta},\mathcal{H}^{BS}_{\theta})$ fully written in a symplectic chart, we can tackle its quantization. The Laplace-Beltrami operator \eqref{laplace-beltrami} is now given by
\begin{align}
K^{BS}_\theta\left(\underline{q},\underline{p}\right)\mapsto&-\frac{\hbar^2}{2}{\Delta^{BS}_{\theta}}_{\underline{q}}=\nonumber\\
&-\frac{\hbar^2}{2}\left(\underline{q}+\theta\underline{q} f\left(\underline{q}\right)\right)\partial_{\underline{q}}\left[\left(\underline{q}+\theta\underline{q} f\left(\underline{q}\right)\right)\partial_{\underline{q}}\right].
\end{align}
This is, of course, the noncommutative counterpart of \eqref{laplace-beltrami-BS}.
Therefore, the noncommutative quantum model is defined by the noncommutative Schr\"odinger equation
\begin{align}
i\partial_{t}\psi\left(\underline{q},t\right)=&\left\{-\frac{\hbar^2}{2}\left(\underline{q}+\theta\underline{q} f\left(\underline{q}\right)\right)\partial_{\underline{q}}\left[\left(\underline{q}+\theta\underline{q} f\left(\underline{q}\right)\right)\partial_{\underline{q}}\right]\right.\nonumber\\
&\left.+U\left(\underline{q}\right)\right\}\psi\left(\underline{q},t\right)\label{nc-schr-BS1a}
\end{align}
As expected, the above quantum-mechanical model formally reduces to the original one \eqref{schr-BS1a} upon taking $\theta\to0$. 

Now, the Wick-rotated version of \eqref{nc-schr-BS1a},
constitutes (after taking $\hbar=1$, appropriately fixing $U\left(\underline{q}\right)$, and assuming the obviated mass term to be $m=1/\sigma^{2}$) a noncommutative quantum mechanics generalization of the standard BS model arising from the classical system \eqref{lag-1}. 
\subsection{Noncommutative extension based on the second approach for a quantum-mechanical representation of the BS model}
The Hamiltonian \eqref{class-ham2a} is written in coordintes $\underline{x}$ as 
\begin{align}
\tilde{\mathcal{H}}^{BS}\left(\underline{q},\underline{p}\right)=&\frac{1}{2}\underline{q}^2\left(\underline{p}+\theta\underline{p}f\left(\underline{q}\right)\right)^2-\alpha\underline{q}\left(\underline{p}+\theta\underline{p}f\left(\underline{q}\right)\right)\nonumber\\
&+U(\underline{q})+\frac{\alpha^2}{2}.
\end{align}
And so, as before, we have ${g_\theta}^{11}$ given by Eq. \eqref{metric-nc-BS} 
serving as the (inverse) metric field for the underlying configuration manifold. Hence, the associated Laplace-Beltrami operator is the same as in the previous case. However, here we must also take care of the term linear in the momentum $\underline{p}$. We therefore get the following associated noncommutative Schr\"odinger equation,
\begin{align}
i\partial_{t}\psi\left(\underline{q},t\right)=&\left\{-\frac{\hbar^2}{2}\left(\underline{q}+\theta\underline{q} f\left(\underline{q}\right)\right)\partial_{\underline{q}}\left[\left(\underline{q}+\theta\underline{q} f\left(\underline{q}\right)\right)\partial_{\underline{q}}\right]\right.\nonumber\\
&\left.-\alpha\underline{q}\left(-i\hbar\partial_{\underline{q}}-i\hbar f\left(\underline{q}\right)\theta\partial_{\underline{q}}\right)
+U\left(\underline{q}\right)+\frac{\alpha^2}{2}\right\}\psi\left(\underline{q},t\right),\label{nc-schr-BS2a}
\end{align}
where we have used the usual prescription $\bar{p}\mapsto-i\hbar\partial_{\underline{q}}$ in order to promote to a differential operator the velocity-dependent term featured in the potential energy function. We have also considered the simplest choice of factor ordering for the term $\underline{p}f\left(\underline{q}\right)$ (as was the case for the commutative counterpart). We note that, as expected, the quantum model defined by \eqref{nc-schr-BS2a} reduces (for $\theta\to 0$) to the commutative one defined by \eqref{schr-BS-2a}. We also stress that for both $\theta\to0$ and $\alpha\to0$ the model directly leads to \eqref{schr-BS1a}. 

\section{Two degrees of freedom model}
Taking advantage of the experience gained, we consider now a system consisting of two degrees of freedom, which itself leads (upon quantization and further Wick rotation) to a model which can be seen as related to a particular subset of the Merton-Garman family \cite{merton, garman}. We also take the opportunity to present two kinds of noncommutative generalizations, the first one directly related to the one implemented on the BS model, and an additional one which is rather popular for systems with several degrees of freedom.

The Merton-Garman models are governed by the equation,
\begin{align}
rC-rSC_S=C_t+\frac{1}{2}VS^2C_{SS}+(\lambda+\mu V)C_V\nonumber\\
+\rho\xi V^{1/(2+\alpha)}SC_{SV} 
+\xi^2V^{2\alpha}C_{VV},\label{merton-garman}    
\end{align}
where $\alpha$, $\lambda$, $\mu$, $\rho$ and $\xi$ are real parameters. The main feature of this pricing model family is that the volatility-related variable $V\equiv\sigma^2$ is assumed to evolve according to a certain Ito process, and it is therefore a stochastic quantity. Specifically, the MG family follows from implementing the Ito calculus to the system 
\begin{align}
dS=rSdt+\sqrt{V}SdW\\
dV=\kappa(\theta-V)dt+\xi V^\alpha d\bar{W}
\end{align}
where $\lambda=\kappa\theta$ and $\mu=-\kappa$; and $W$, $\bar{W}$ are Wiener processes with correlation $\rho\in[-1,1]$ (see, e.g. \cite{baaquie2}). At first sight, the underlying stock price appears to evolve in a manner similar to that of the BS model, but notice that now the (squared) volatility $V$ is assumed to evolve according to the Ito process defined by the second equation of the system above.

Consider a mechanical system whose configuration manifold $(\mathcal{Q}^{MG},g^{MG})$ is coordinatized by the pair $(q^1,q^2)=(q,w)$ where $q,w$ are strictly positive, and in which the components of the (inverse) metric field $g$ are given by 
\begin{equation}
{g^{MG}}^{11}=q^2w,\quad {g^{MG}}^{22}=2\xi^2w^2,\quad {g^{MG}}^{12}=0=g^{21}. \label{metric-MG}
\end{equation}
Assume hence a Lagrangian 
\begin{equation}
\mathcal{L}^{MG}(q,w,\dot{q},\dot{w})=\frac{1}{2}
\left[\frac{1}{w}\left(\frac{\dot{q}}{q}\right)^2
+\frac{1}{2\xi^2}\left(\frac{\dot{w}}{w}\right)^2\right]-U(q,w).\label{lag-MG}
\end{equation}
The associated Hamiltonian is therefore
\begin{align}
\mathcal{H}^{MG}(q,w,p,k)&=\frac{1}{2}\left(q^2wp^2+2\xi^2w^2k^2\right)+U(q,w)\label{ham-MG}\\
&=K^{MG}(q,w,p,k)+U(q,w).
\nonumber
\end{align}
The symplectic structure of the cotangent bundle $\Gamma^{MG}$ of the configuration space $(\mathcal{Q}^{MG},g^{MG})$ can then be written (in the canonical coordinates $x=(q,w,p,k)$) as
\begin{equation}
{\omega^{MG}}_x=dp\wedge dq+dk\wedge dw.
\end{equation}

Now, the corresponding Laplace-Beltrami operator is given by 
\begin{align}
K^{MG}(q,w,p,k)\mapsto-\frac{\hbar^2}{2}\Delta^{MG}_{q,w}=\nonumber\\ 
\Delta^{MG}_{q,w}=
\xi qw^{3/2}\left[\partial_q\left(\frac{q^2w}{\xi qw^{3/2}}\partial_q\right)+\partial_w\left(\frac{2\xi^2w^2}{\xi qw^{3/2}}\partial_w\right)\right]
\end{align}
and so, the associated Schr\"odinger equation is 
\begin{align}
i\partial_t\psi(q,w,t)=\left\{-\frac{\hbar^2}{2}\xi qw^{3/2}\left[\partial_q\left(\frac{q^2w}{\xi qw^{3/2}}\partial_q\right)+\partial_w\left(\frac{2\xi^2w^2}{\xi qw^{3/2}}\partial_w\right)\right]\right.\nonumber\\
\left.+U(q,w)\right\}\psi(q,w,t).\label{schrodinger-MG}
\end{align}
By performing a Wick rotation, we obtain a model related to the particular subset of the family \eqref{merton-garman} associated to $\alpha=1$, $\lambda=0$, $\mu=2\xi^2$, $\rho=0$. The main difference being that in the obtained equation, a factor of $w$ (which in this case is identified with the volatiliy-related variable $V$) is featured in the term linear in $\partial_q$.

\subsection{A noncommutative quantum mechanics generalization}
In order to construct a simple noncommutative generalization we consider first a quantum algebra of the type \eqref{nc-algebra}, i.e,  
\begin{align}
\{q,p\}=1+\theta f(q)\longleftrightarrow
[\hat{q},\hat{p}]=i\hbar\left(\hat{1}+\theta\hat{f}(\hat{q})\right);\\ 
\{w,k\}=1+\theta g(w)\longleftrightarrow
[\hat{w},\hat{k}]=i\hbar\left(\hat{1}+\theta\hat{g}(\hat{w})\right), \label{nc-algebra-MG}
\end{align} 
with the remaining commutation relations being the usual trivial ones. The Poisson structure $\pi_\theta^{MG}$ of the classical phase space can then be written as
\begin{equation}
{\pi_{\theta}^{MG}}_x=\left(1+\theta f(q)\right)\partial_{q}\wedge\partial_{p}+\left(1+\theta g(w)\right)\partial_{w}\wedge\partial_{k}\label{nc-pois-structure-mg},
\end{equation}
while the associated symplectic structure $\omega_{\theta}^{MG}$  is given by, 
\begin{equation}
{\omega_{\theta}^{MG}}_x=\left(1+\theta f(q)\right)^{-1}dp\wedge dq+\left(1+\theta g(w)\right)^{-1}dk\wedge dw \label{nc-symp-structure-mg}.
\end{equation}
Classically, we therefore have a mechanical system $(\Gamma_\theta^{MG},\mathcal{H}^{MG})$, where $\Gamma_\theta^{MG}$ is the cotangent bundle of some underlying configuration manifold $(\mathcal{Q}^{MG}_\theta,g^{MG}_\theta)$, and where the symplectic structure is given by \eqref{nc-symp-structure-mg} (with the same Hamiltonian function \eqref{ham-MG} of the original mechanical system $(\Gamma^{MG},\mathcal{H}^{MG})$).

By going to a symplectic patch $\underline{x}=\left(\underline{q},\underline{w},\underline{p},\underline{k}\right)$ defined by 
\begin{equation}
q=\underline{q},\quad w=\underline{w},\quad p=\underline{p}+\theta\underline{p}f(\underline{q}),\quad k=\underline{k}+\theta\underline{k}g(\underline{w}),
\end{equation}
the Hamiltonian \eqref{ham-MG} is written as 
\begin{align}
\mathcal{H}^{MG}\left(\underline{q},\underline{w},\underline{p},\underline{k}\right)&=\frac{1}{2}\left(\underline{q}^2\underline{w}\left(\underline{p}+\theta\underline{p}f(\underline{q})\right)^2+2\xi^2\underline{w}^2\left(\underline{k}+\theta\underline{k}g(\underline{w})\right)^2\right)\nonumber\\
+U\left(\underline{q},\underline{w}\right)\\&=K^{MG}_\theta\left(\underline{q},\underline{w},\underline{p},\underline{k}\right)+U\left(\underline{q},\underline{w}\right).\nonumber\label{def-ham-MG}
\end{align}
Notice that the (inverse) metric field $g_\theta^{MG}$ of the underlying configuration manifold $(\mathcal{Q}_\theta^{MG},g_\theta^{MG})$ is given by 
\begin{equation}
{g_\theta^{MG}}^{11}=w\left(q+\theta qf\left(\underline{q}\right)\right)^2,\,   {g_\theta^{MG}}^{22}=2\xi^2\left(w+\theta wf\left(\underline{w}\right)\right)^2,\label{metric-nc-MG1} 
\end{equation}
the remaining components being null.

The quantum counterpart is achieved by constructing the associated Laplace-Beltrami operator, which is given by, 
\begin{align}
K^{MG}_\theta\left(\underline{q},\underline{w},\underline{p},\underline{k}\right)
\mapsto
-\frac{\hbar^2}{2}{\Delta^{MG}_\theta}_{\underline{q},\underline{w}}\nonumber\\ 
{\Delta^{MG}_\theta}_{\underline{q},\underline{w}}=
\xi\sqrt{2\underline{w}}\left(\underline{q}+\theta\underline{q}f(\underline{q})\right)\left(\underline{w}+\theta \underline{w}g(\underline{w})\right)\nonumber\\
\times\left[
\partial_{\underline{q}}
\left(\frac{\underline{w}\left(\underline{q}+\theta\underline{q}f(\underline{q})\right)^2}{\xi\sqrt{2\underline{w}}\left(\underline{q}+\theta\underline{q} f(\underline{q})\right)\left(\underline{w}+\theta\underline{w} g(\underline{w})\right)}\partial_{\underline{q}}\right)\nonumber\right.\\
\left.+
\partial_{\underline{w}}
\left(\frac{2\xi^2\left(\underline{w}+\theta\underline{w}g(\underline{w})\right)^2}{\xi\sqrt{2\underline{w}}\left(\underline{q}+\theta\underline{q} f(\underline{q})\right)\left(\underline{w}+\theta\underline{w} g(\underline{w})\right)}\partial_{\underline{w}}\right)
\right]
\end{align}
leading to the rather formidable Schr\"odinger equation,
\begin{align}
i\partial_t\psi\left(\underline{q},\underline{w},t\right)=\left\{-\frac{\hbar^2}{2}
\xi\sqrt{2\underline{w}}\left(\underline{q}+\theta\underline{q}f(\underline{q})\right)\left(\underline{w}+\theta \underline{w}g(\underline{w})\right)\right.\nonumber\\
\times\left.\left[
\partial_{\underline{q}}
\left(\frac{\underline{w}\left(\underline{q}+\theta\underline{q}f(\underline{q})\right)^2}{\xi\sqrt{2\underline{w}}\left(\underline{q}+\theta\underline{q} f(\underline{q})\right)\left(\underline{w}+\theta\underline{w} g(\underline{w})\right)}\partial_{\underline{q}}\right)\nonumber\right.\right.\\
\left.\left.+
\partial_{\underline{w}}
\left(\frac{2\xi^2\left(\underline{w}+\theta\underline{w}g(\underline{w})\right)^2}{\xi\sqrt{2\underline{w}}\left(\underline{q}+\theta\underline{q} f(\underline{q})\right)\left(\underline{w}+\theta\underline{w} g(\underline{w})\right)}\partial_{\underline{w}}\right)
\right]\right.\nonumber\\
\left.+U\left(\underline{q},\underline{w}\right)\right\}\psi\left(\underline{q},\underline{w},t\right).\label{nc1-schrodinger-MG}
\end{align}
We note that upon taking $\theta\to0$ on the above equation we recover the quantum model defined by the Schr\"odinger equation \eqref{schrodinger-MG}, as expected.

\subsection{Another noncommutative generalization}
We turn now to a different noncommutative structure, which in the noncommutative quantum mechanics paradigm is highly relevant for the case of several degrees of freedom. Consider the  noncommutative quantum algebra ($\eta$ a real constant)
\begin{align}
\{p,k\}=\eta\longleftrightarrow [\hat{p},\hat{k}]=i\eta,\\
\{q,p\}=1=\{w,k\}\longleftrightarrow[\hat{q},\hat{p}]=i\hbar=[\hat{w},\hat{k}]\nonumber\label{nc-algebra-MG2}
\end{align}
(with the commutation relation among the configuration variables $q$ and $w$ being the usual trivial one). 

This particular type of noncommutativity enjoys of a certain degree of popularity among the literature related to noncommutative quantum mechanics (see, e.g. \cite{ncqm-review}). 

Classically, we are then dealing with a mechanical system $(\Gamma_\eta^{MG},\mathcal{H}^{MG})$ in which the Poisson structure $\pi^{MG}_\eta$ can be written in coordinates $x=(q,w,p,k)$ as 
\begin{equation}
{\pi^{MG}_\eta}_x=\partial_q\wedge\partial_p+\partial_w\wedge\partial_k+\eta\partial_p\wedge\partial_k,
\end{equation}
while the associated symplectic structure $\omega^{MG}_\eta$ is now given by
\begin{equation}
{\omega^{MG}_\eta}_x=dp\wedge dq+dk\wedge dw+\eta dq\wedge dw,
\end{equation}
and where the Hamiltonian function is \eqref{ham-MG}. 

As in the previous noncommutative cases, we seek canonical coordinates in order to carry out the quantization of the model in a standard way. It is straightforward to verify that one such symplectic patch is defined by
\begin{equation}
q=\underline{q},\quad
w=\underline{w},\quad 
p=\underline{p}+\eta\underline{w},\quad
k=\underline{k}
\end{equation}
In these new canonical variables, the Hamiltonian \eqref{ham-MG} is written as
\begin{align}
\mathcal{H}^{MG}\left(\underline{q},\underline{w},\underline{p},\underline{k}\right)
&=\frac{1}{2}\left(\underline{q}^2\underline{w}\left(\underline{p}+\eta \underline{w}\right)^2+2\xi^2\underline{w}^2\underline{k}^2\right)+U\left(\underline{q},\underline{w}\right)\label{def-ham-MG2}\\
&=\frac{1}{2}\left(\underline{q}^2\underline{w}\underline{p}^2+2\xi^2\underline{w}^2\underline{k}^2\right)+U\left(\underline{q},\underline{w}\right)\nonumber\\
+\eta\underline{q}^2\underline{w}^2\underline{p}+\frac{1}{2}\eta\underline{q}^2\underline{w}^3\nonumber\\
&=K^{MG}_\eta\left(\underline{q},\underline{w},\underline{p},\underline{k}\right)+U_\eta\left(\underline{q},\underline{w},\underline{p}\right).\nonumber
\end{align}
We observe that the (inverse) metric field of the underlying configuration manifold $(\mathcal Q^{MG}_\eta,g^{MG}_\eta)$ is given by 
\begin{equation}
{g^{MG}_\eta}^{11}=\underline{q}^2\underline{w},\quad
{g^{MG}_\eta}^{22}=2\xi^2\underline{w}^2,\quad 
{g^{MG}_\eta}^{12}=0={g^{MG}_\eta}^{21}, \label{metric-nc-MG2}
\end{equation}
and so the associated Laplace-Beltrami operator is given by
\begin{align}
K^{MG}_\eta\left(\underline{q},\underline{w},\underline{p},\underline{k}\right)\mapsto-\frac{\hbar^2}{2}{\Delta^{MG}_\eta}_{\underline{q},\underline{w}}\nonumber\\ 
{\Delta^{MG}_\eta}_{\underline{q},\underline{w}}=\xi \underline{q}\underline{w}^{3/2}\left[\partial_{\underline{q}}\left(\frac{\underline{q}^2\underline{w}}{\xi \underline{q}\underline{w}^{3/2}}\partial_{\underline{q}}\right)+\partial_{\underline{w}}\left(\frac{2\xi^2\underline{w}^2}{\xi \underline{q}\underline{w}^{3/2}}\partial_{\underline{w}}\right)\right]
\end{align}
However, the deformed Hamiltonian \eqref{def-ham-MG2} now features additional terms in the potential function, one of them being linear in $\underline{p}$. By using the usual identification $\underline{p}\mapsto-i\hbar\partial_{\underline{q}}$ in order to promote this term to a differential operator, we get the following Schr\"odinger equation, 
\begin{align}
i\partial_t\psi\left(\underline{q},\underline{w},t\right)=\left\{-\frac{\hbar^2}{2}\xi\underline{q}\underline{w}^{3/2}\left[\partial_{\underline{q}}\left(\frac{\underline{q}^2\underline{w}}{\xi \underline{q}\underline{w}^{3/2}}\partial_{\underline{q}}\right)+\partial_{\underline{w}}\left(\frac{2\xi^2\underline{w}^2}{\xi \underline{q}\underline{w}^{3/2}}\partial_{\underline{w}}\right)\right]\right.\nonumber\\
\left.+\eta\underline{q}^2\underline{w}^2(-i\hbar\partial_{\underline{q}})+\frac{1}{2}\eta \underline{q}^2\underline{w}^3+U\left(\underline{q},\underline{w}\right)\right\}\psi(q,w,t).\label{nc-schrodinger-MG2}
\end{align}
This noncommutative Schr\"odinger equation reduces to the commutative one \eqref{schrodinger-MG} upon taking $\eta\to0$. Observe that (the Wick-rotated version of) the above equation constitutes a more gentle departure from (the Wick-rotated version of) \eqref{schrodinger-MG} than the deviation provided by (the Wick-rotated version of) \eqref{nc1-schrodinger-MG}, the only difference being that two additional terms are now featured in the potential operator.


\section{Summary and final remarks}

Let us summarize the most significant findings of the present investigation; highlight and briefly discuss what we believe are some important points (in italics); and state possible trends  that can be followed within the framework presented in the present letter (in italics).
\begin{enumerate}
%
\item What we have called the first approach for a quantum-mechanical representation of the BS model was framed to the quantization of a mechanical model whose Hamiltonian is
\begin{equation*}
\mathcal{H}^{BS}(q,p)=\frac{1}{2}q^2p^2+U(q).
\end{equation*}
The associated Laplace-Beltrami operator achieving a well-defined quantization being 
\begin{equation*}
K^{BS}(q,p)\mapsto-\frac{\hbar^2}{2}\Delta^{BS}_q=-\frac{\hbar^2}{2}\left(q^2\partial^2_q+q\partial_q\right).
\end{equation*}
The resulting Schr\"odinger equation is
\begin{equation*}
i\partial_t\psi(q,t)=\left[-\frac{\hbar^2}{2}\left(q^2\partial^2_q+q\partial_q\right)+U(q)\right]\psi(q,t).
\end{equation*} 
With respect to the second route, we showed that it can be understood as the quantization of a mechanical model whose Hamiltonian is 
\begin{equation*}
\tilde{\mathcal{H}}^{BS}(q,p)=\frac{1}{2}q^2p^2-\alpha qp+U(q)+\frac{\alpha^2}{2}.
\end{equation*}
The associated Laplace-Beltrami operator achieving a well-defined quantization being
\begin{equation*}
\tilde{K}^{BS}(q,p)\mapsto-\frac{\hbar^2}{2}{\Delta}^{BS}_q=-\frac{\hbar^2}{2}\left(q^2\partial_q^2+q\partial_q\right),
\end{equation*}
The resulting Schr\"odinger equation is
\begin{align*}
i\partial_t\psi(q,t)=\left[-\frac{\hbar^2}{2}\left(q^2\partial_q^2+q\partial_q\right)-\alpha q(-i\hbar\partial_q)\right.\nonumber\\
\left.+U(q)+\frac{\alpha^2}{2}\right]\psi(q,t),
\end{align*}
where the term linear in momentum $p$ was promoted to a differential operator via the usual recipee.
\begin{itemize}
\item\emph{Two well defined quantum mechanical representations of the BS model were obtained, starting from a Lagrangian formulation of a classical system and then carrying out the quantization via the associated Laplace-Beltrami operator. This is in contrast with the usual treatment given so far in the standard literature, where a Lagrangian is obtained only a posteriori, and where the quantization is carried by means of the naive rule $p\mapsto-i\hbar\partial_q$ (which is known to lead to incorrect quantization schemes in cases where the configuration manifold is not flat and/or the used configuration variables are not cartesian). We note that the classical framework put forward here is not the same as the one obtained in the literature.}
\item\emph{The constructed Laplace-Beltrami operators (one for each model) are essentially the unique ones which lead to the BS model, since the second derivative term of the BS model fixes completely the structure of the Laplace-Beltrami operator that should be considered. This in turn completely fixes the whole geometric structure of the associated classical framework, which was clearly recognized and emphasized. This in turn prompts the use of global geometric techniques for the further study and generalization of such models.}
\item\emph{We stress that the obtained quantum models are not equivalent (even when both represent the BS model), in the sense that they arise from two non equivalent classical systems.}
\item\emph{The semiclassical scheme provided by the associated path integral quantization would not turn out to be the same as the one reported in the existing literature, where the Lagrangian is obtained indirectly by a kind of inverse problem process from a quantum framework based on the naive quantization rule $p\mapsto-i\hbar\partial_q$ (see, e.g. \cite{baaquie2} and \cite{contreras}).}
\end{itemize}
\item An extension of the BS model to the realm of the so called noncommutative quantum mechanics was put forward. The aforementioned geometrical structure was instrumental in making such generalization. 

More specifically, the noncommutative algebra 
\begin{equation*}
\{q,p\}=1+\theta f(q)\longleftrightarrow [\hat{q},\hat{p}]=i\hbar\left(\hat{1}+\theta\hat{f}(\hat{q})\right)
\end{equation*}
was considered, whose classical limit features a phase space with a symplectic structure 
\begin{equation*}
{\omega_{\theta}^{BS}}_{x}=\left(1+\theta f(q)\right)^{-1}dp\wedge dq.
\end{equation*}
Quantization is achieved by going to a symplectic chart, in which the Hamiltonian for the first approach to a quantum-mechanical interpretation of the BS model takes the form 
\begin{equation*}
\mathcal{H}^{BS}\left(\underline{q},\underline{p}\right)=\frac{1}{2}\underline{q}^2\left[\underline{p}+\theta\underline{p} f\left(\underline{q}\right)\right]^2+U\left(\underline{q}\right).
\end{equation*}
The associated Laplace-Beltrami operator results to be 
\begin{align*}
K^{BS}_\theta\left(\underline{q},\underline{p}\right)\mapsto&-\frac{\hbar^2}{2}{\Delta^{BS}_{\theta}}_{\underline{q}}=\nonumber\\
&-\frac{\hbar^2}{2}\left(\underline{q}+\theta\underline{q} f\left(\underline{q}\right)\right)\partial_{\underline{q}}\left[\left(\underline{q}+\theta\underline{q} f\left(\underline{q}\right)\right)\partial_{\underline{q}}\right].
\end{align*}
The corresponding Schr\"odinger equation is 
\begin{align*}
i\partial_{t}\psi\left(\underline{q},t\right)=&\left\{-\frac{\hbar^2}{2}\left(\underline{q}+\theta\underline{q} f\left(\underline{q}\right)\right)\partial_{\underline{q}}\left[\left(\underline{q}+\theta\underline{q} f\left(\underline{q}\right)\right)\partial_{\underline{q}}\right]\right.\nonumber\\
&\left.+U\left(\underline{q}\right)\right\}\psi\left(\underline{q},t\right).
\end{align*}
In the case of the second avenue, the Hamiltonian function takes the form 
\begin{align*}
\tilde{\mathcal{H}}^{BS}\left(\underline{q},\underline{p}\right)=&\frac{1}{2}\underline{q}^2\left(\underline{p}+\theta\underline{p}f\left(\underline{q}\right)\right)^2-\alpha\underline{q}\left(\underline{p}+\theta\underline{p}f\left(\underline{q}\right)\right)\nonumber\\
&+U(\underline{q})+\frac{\alpha^2}{2}.
\end{align*}
The Laplace-Beltrami operator being the same as for the first approach. The Associated Schr\"odinger equation is 
\begin{align*}
i\partial_{t}\psi\left(\underline{q},t\right)=&\left\{-\frac{\hbar^2}{2}\left(\underline{q}+\theta\underline{q} f\left(\underline{q}\right)\right)\partial_{\underline{q}}\left[\left(\underline{q}+\theta\underline{q} f\left(\underline{q}\right)\right)\partial_{\underline{q}}\right]\right.\nonumber\\
&\left.-\alpha\underline{q}\left(-i\hbar\partial_{\underline{q}}-i\hbar f\left(\underline{q}\right)\theta\partial_{\underline{q}}\right)
+U\left(\underline{q}\right)+\frac{\alpha^2}{2}\right\}\psi\left(\underline{q},t\right).
\end{align*}
\begin{itemize}
\item \emph{Here we should remark that a more general noncommutativity structure can ben considered, namely that of \eqref{nc-general-algebra}. However, in such more general case, achieving a quantization akin to the standard commutative one would be rather involved, since terms cubic and quartic in the momentum would be featured in the classical Hamiltonian when expressed in a symplectic patch.}
\item \emph{Due to the generic function $f(q)$ featured in the extended model, the noncommutative generalization of the BS model presented here would be able to accommodate additional potential-like terms, which could in turn be used to describe financial developments not accounted for in the standard BS model. Studies of some particular deviations giving rise to new meaningful financial issues are therefore a natural pursuit. One such study is currently being tackled by the authors and will be reported elsewhere as a follow up to the present letter.}
\end{itemize}
\item A classical system incorporating two degrees of freedom was also considered, the quantization of which led to a model related to the MG family. This model was also generalized via the noncommutative quantum mechanics paradigm. 

The classical system considered is defined by the Hamiltonian 
\begin{align*}
\mathcal{H}^{MG}(q,w,p,k)&=\frac{1}{2}\left(q^2wp^2+2\xi^2w^2k^2\right)+U(q,w).
\nonumber
\end{align*}
The associated Laplace-Beltrami operator being 
\begin{align*}
K^{MG}(q,w,p,k)\mapsto-\frac{\hbar^2}{2}\Delta^{MG}_{q,w},\nonumber\\ 
\Delta^{MG}_{q,w}=
\xi qw^{3/2}\left[\partial_q\left(\frac{q^2w}{\xi qw^{3/2}}\partial_q\right)+\partial_w\left(\frac{2\xi^2w^2}{\xi qw^{3/2}}\partial_w\right)\right],\nonumber
\end{align*}
while the corresponding Schr\"odinger equation is \eqref{schrodinger-MG}.

The noncoommutative generalization was carried considering two schemes. The first one is defined by the noncommutative structure 
\begin{align*}
\{q,p\}=1+\theta f(q)\longleftrightarrow
[\hat{q},\hat{p}]=i\hbar\left(\hat{1}+\theta\hat{f}(\hat{q})\right);\\ 
\{w,k\}=1+\theta g(w)\longleftrightarrow
[\hat{w},\hat{k}]=i\hbar\left(\hat{1}+\theta\hat{g}(\hat{w})\right), 
\end{align*}
leading to a classical phase space with symplectic structure
\begin{equation*}
{\omega_{\theta}^{MG}}_x=\left(1+\theta f(q)\right)^{-1}dp\wedge dq+\left(1+\theta g(w)\right)^{-1}dk\wedge dw.
\end{equation*}
By going to a symplectic chart, the Hamiltonian is written as 
\begin{align*}
\mathcal{H}^{MG}\left(\underline{q},\underline{w},\underline{p},\underline{k}\right)=\frac{1}{2}\left(\underline{q}^2\underline{w}\left(\underline{p}+\theta\underline{p}f(\underline{q})\right)^2\nonumber\right.\\\left.+2\xi^2\underline{w}^2\left(\underline{k}+\theta\underline{k}g(\underline{w})\right)^2\right)
+U\left(\underline{q},\underline{w}\right).
\end{align*}
The associated Laplace-Beltrami operator is given by
\begin{align*}
K^{MG}_\theta\left(\underline{q},\underline{w},\underline{p},\underline{k}\right)
\mapsto
-\frac{\hbar^2}{2}{\Delta^{MG}_\theta}_{\underline{q},\underline{w}}\nonumber\\ 
{\Delta^{MG}_\theta}_{\underline{q},\underline{w}}=
\xi\sqrt{2\underline{w}}\left(\underline{q}+\theta\underline{q}f(\underline{q})\right)\left(\underline{w}+\theta \underline{w}g(\underline{w})\right)\nonumber\\
\times\left[
\partial_{\underline{q}}
\left(\frac{\underline{w}\left(\underline{q}+\theta\underline{q}f(\underline{q})\right)^2}{\xi\sqrt{2\underline{w}}\left(\underline{q}+\theta\underline{q} f(\underline{q})\right)\left(\underline{w}+\theta\underline{w} g(\underline{w})\right)}\partial_{\underline{q}}\right)\nonumber\right.\\
\left.+
\partial_{\underline{w}}
\left(\frac{2\xi^2\left(\underline{w}+\theta\underline{w}g(\underline{w})\right)^2}{\xi\sqrt{2\underline{w}}\left(\underline{q}+\theta\underline{q} f(\underline{q})\right)\left(\underline{w}+\theta\underline{w} g(\underline{w})\right)}\partial_{\underline{w}}\right)
\right],
\end{align*}
which in turn yields the rather cumbersome schr\"odinger equation \eqref{nc1-schrodinger-MG}.

The second noncommutative structure which was considered is defined by the algebra 
\begin{align*}
\{p,k\}=\eta\longleftrightarrow [\hat{p},\hat{k}]=i\eta,\\
\{q,p\}=1=\{w,k\}\longleftrightarrow[\hat{q},\hat{p}]=i\hbar=[\hat{w},\hat{k}],\nonumber
\end{align*}
which amounts to a classical phase space with symplectic structure 
\begin{equation*}
{\omega^{MG}_\eta}_x=dp\wedge dq+dk\wedge dw+\eta dq\wedge dw.
\end{equation*}
Using canonical coordinates, the Hamiltonian function is written as 
\begin{align*}
\mathcal{H}^{MG}\left(\underline{q},\underline{w},\underline{p},\underline{k}\right)
&=\frac{1}{2}\left(\underline{q}^2\underline{w}\left(\underline{p}+\eta \underline{w}\right)^2+2\xi^2\underline{w}^2\underline{k}^2\right)\nonumber\\
+U\left(\underline{q},\underline{w}\right).\label{def-ham-MG2}
\end{align*}
And so, the associated Laplace-Beltrami operator is 
\begin{align*}
K^{MG}_\eta\left(\underline{q},\underline{w},\underline{p},\underline{k}\right)\mapsto-\frac{\hbar^2}{2}{\Delta^{MG}_\eta}_{\underline{q},\underline{w}}\nonumber\\ 
{\Delta^{MG}_\eta}_{\underline{q},\underline{w}}=\xi \underline{q}\underline{w}^{3/2}\left[\partial_{\underline{q}}\left(\frac{\underline{q}^2\underline{w}}{\xi \underline{q}\underline{w}^{3/2}}\partial_{\underline{q}}\right)+\partial_{\underline{w}}\left(\frac{2\xi^2\underline{w}^2}{\xi \underline{q}\underline{w}^{3/2}}\partial_{\underline{w}}\right)\right].
\end{align*}
The corresponding Schr\"odinger equation is \eqref{nc-schrodinger-MG2}.

\begin{itemize}
\item\emph{We note that the commutative MG-related model features a factor of $w$ (which is to be identified with the volatility-related variable $V$ of the MG family) as part of the term linear in $\partial_q$. This would amount to generalize the MG model to one having an effective, time-dependent interest rate.}
\item\emph{As was the case of the BS model, the first noncommutative generalization obtained features terms involving generic functions $f(q)$ and $g(w)$, which could account for meaningful financial developments. The second noncommutative model presents a less radical departure from the commutative one, and features a velocity-dependent potential term, as well as a polynomial one.}
\item\emph{The above observations pave the way for the engagement in dedicated studies for the presented MG-related model, and for the two given noncommutative generalizations (the first one being evidently more easy to tackle).}
\end{itemize}
%
%

\end{enumerate}

\section*{Acknowledgments}
A. E. G. was partially supported by SNI-CONAHCyT; L. R. D. B. was partially supported by SNI-CONAHCyT. 
		

\end{document}